\documentstyle[preprint,tighten,aps,epsfig]{revtex}
\begin{document}
\draft

\title{Radiative decays of decuplet hyperons}
\author{Georg Wagner$^{1,2}$\footnote{supported by a postdoctoral 
          fellowship of the Deutsche Forschungsgemeinschaft (DFG)
          under contract number Wa1147/1-1. },
        A.\ J.\ Buchmann$^2$, and Amand Faessler$^2$}
\address{$^1$ Centre for the Subatomic Structure of Matter (CSSM),  \\
         University of Adelaide, Australia 5005}
\address{$^2$ Institut f\"{u}r Theor.\ Physik, 
         Universit\"{a}t T\"{u}bingen,  \\
         Auf der Morgenstelle 14, 72076 T\"{u}bingen, Germany}
\date{\today }        

\maketitle

\begin{abstract} 
  We calculate the radiative decay widths of decuplet
  hyperons in a chiral constituent quark model including 
  electromagnetic exchange currents between quarks.
  Exchange currents contribute significantly to the E2 transition amplitude,
  while they largely cancel for the M1 transition amplitude.
  Strangeness suppression of the radiative hyperon decays
  is found to be weakened by exchange currents.
  Differences and similarities between our results and 
  other recent model predictions are discussed.
\end{abstract}

 \def\shiftleft#1{#1\llap{#1\hskip 0.04em}}
 \def\shiftdown#1{#1\llap{\lower.04ex\hbox{#1}}}
 \def\thick#1{\shiftdown{\shiftleft{#1}}}    
 \def\b#1{\thick{\hbox{$#1$}}}    
\section{Introduction}
Electromagnetic transitions in baryons provide not only information on the 
importance of exchange currents but also on the effective quark-quark 
interaction.
In the context of potential models, electromagnetic gauge invariance
relates the two-body terms of the quark model Hamiltonian (potentials) 
to the two-body terms in the current operator (exchange currents). 
The excitation spectrum and electromagnetic transition
amplitudes of baryons are thus intimately connected.
Recently, several works \cite{Buc91,Rob93,Wag95,Buc97,Mey97} systematically 
discuss two-body exchange  current effects on electromagnetic   
observables.
A good example for the importance of exchange currents is the C2 (E2) multipole
amplitude in the $\gamma N\leftrightarrow \Delta$ transition.
While constituent quark model calculations using D-state admixtures 
underpredict this observable by a factor of three or more, exchange currents 
give the correct empirical quadrupole transition amplitude \cite{Buc97}.

Here, we briefly report on the first study of exchange current effects on 
the radiative decays of all decuplet hyperons.
Theoretical studies of radiative hyperon decays have been performed in 
the pioneering work of Lipkin \cite{Lip73},
the quark model (without exchange currents) \cite{Dar83},
SU$_F$(3) Skyrme model approaches \cite{Aba96,Oh95,Sch95},
chiral bag models \cite{Kum88},
heavy baryon chiral perturbation theory \cite{But93}, or
quenched lattice calculations \cite{Lei93}.
Current experimental programs aim at a detailed measurement of the radiative
decays of some $\Sigma^{\ast}$ and $\Xi^{\ast}$ hyperons \cite{Rus95}.

Radiative decays of hyperons are interesting for several reasons.
In previous quark model calculations of decuplet hyperon decays \cite{Dar83}, 
which neglect exchange currents and $D$-state admixtures, all decays are 
pure M1 transitions. Here, we find that the inclusion of 
exchange currents leads in all cases to nonvanishing E2/M1 ratios.
The comparison of our results with other model predictions and experimental 
data may not only provide another signal of exchange currents inside baryons
but may even help to pin down the relative importance of vector (gluon) vs.\ 
pseudoscalar degrees of freedom in the effective quark-quark interaction at 
low energies.

Radiative hyperon decays are sensitive to 
SU$_F$(3) flavor symmetry breaking and strangeness suppression.
The decay widths of the negatively charged hyperons
$\Sigma^{\ast -}\rightarrow\gamma\Sigma^-$ and
$\Xi^{\ast-}\rightarrow\gamma\Xi^-$
would be zero, if SU$_F$(3) flavor-symmetry was realized in nature.
It has been speculated \cite{Aba96} that these two decays remain almost
forbidden even after SU$_F$(3) symmetry breaking.
Strangeness suppression, i.\ e.\ the decrease of the decay amplitude with 
increasing strangeness of the hyperon, is best studied by comparing transitions
involving wave functions which are identical except for the replacement 
of d-quarks by s-quarks.
The decays  $\gamma n\leftrightarrow \Delta^{0}$ and
$\gamma \Xi^{0}\leftrightarrow \Xi^{\ast 0}$ are particularly suited,
because the strangeness content increases by two units.

\section{Model description}
As a consequence of the spontaneously broken chiral symmetry of
low-energy QCD, constituent quarks and the pseudoscalar (PS)
mesons emerge as relevant degrees of freedom in hadron physics.
The chiral quark model Hamiltonian in the case of three non-equal
quark masses $m_i$ is
\begin{equation}
  H = \sum_{i=1}^{3} \big( m_i+ {{\bf p}^2_i\over 2m_i} \big)
  - {{\bf P}^2\over 2M} - a_c \sum_{i<j}^{3}
  \b{\lambda}^C_i\cdot \b{\lambda}^C_j \, ({\bf r }_i-{\bf r }_j)^2
  + \sum_{i<j}^{3} V^{Res}({\bf r}_i,{\bf r}_j)  \; .
\label{eq:ham}
\end{equation}
Here, $\b{\lambda}^C_i \cdot \b{\lambda}^C_j = \sum_{a=1}^{8} \lambda^{C,a}_i 
\lambda^{C,a}_j$ is a scalar product in color space, where
$\lambda^{C,a}_i$ are the Gell-Mann SU$_C$(3) color matrices.
A quadratic confinement potential is used.
The radial form of the confinement potential is according to our experience 
not crucial for the discussion of hadronic ground state properties.
We will discuss the dependencies of our results on different types of 
confinement interactions, e.\ g.\ linear confinement, elsewhere.  
Hamiltonian (\ref{eq:ham}) is described for the two-flavor case in
Refs.\ \cite{Buc91,Wag95,Buc97}.
The residual interactions $V^{Res}$ comprise one-gluon exchange (OGE)
in the common Fermi-Breit form without retardation corrections \cite{deR75}, 
\begin{eqnarray}
  V^{\rm{OGE}} &=&
  {\alpha_{s}\over 4}  \b{\lambda}^C_{i} \!\cdot\!\b{\lambda}^C_{j} 
  \!\biggl\lbrace  {1\over r} - {\pi\over 2}({1\over m_i^2}+{1\over m_j^2} +
  {4\over 3} {\b{\sigma}_{i}\cdot\b{\sigma}_{j}\over m_im_j} ) \,
  \delta({\bf r}) - {1\over 4m_im_j} (3\b{\sigma}_i\cdot{\bf{\hat r}}
  \b{\sigma}_j\cdot{\bf {\hat r}} -\b{\sigma}_i\cdot\b{\sigma}_j) {1\over r^3} 
\nonumber \\ 
  &-& {1\over 8r^3}\!\bigg( 3\big( {\bf r}  \!\times\! 
  \! ( {{\bf p}_i\over m_i} - {{\bf p}_j\over m_j} )\big)\!\cdot\! 
  ( {\b{\sigma}_i\over m_i} + {\b{\sigma}_j\over m_j} ) 
  \! -\! \big( {\bf r}\!\times\!
  \! ( {{\bf p}_i\over m_i} + {{\bf p}_j\over m_j} )\big)\!\cdot\! 
  ( {\b{\sigma}_i\over m_i} - {\b{\sigma}_j\over m_j} ) \bigg) 
  \biggr\rbrace\; ,
\label{eq:gluon}
\end{eqnarray}
and the 
generalized chiral interactions due to pseudoscalar (PS) meson exchange.
\begin{eqnarray}
  V^{\rm{PS}} &=&
  \left( \b{\sigma}_i\cdot \b{\nabla}_{\bf r}\right)
  \left( \b{\sigma}_j\cdot \b{\nabla}_{\bf r}\right)
  \left\{ \sum_{a=1}^{3} \lambda_i^a\cdot\lambda_j^a \tilde{V}_\pi (r) +
          \sum_{a=4}^{7} \lambda_i^a\cdot\lambda_j^a \tilde{V}_K (r) +
                         \lambda_i^8\cdot\lambda_j^8 \tilde{V}_{\eta} (r) 
  \right\}
\nonumber
\\ 
  \tilde{V}_\gamma &=& \frac{g_{{\rm{PS}}qq}^2}{4\pi} 
  \frac{\Lambda_\gamma^2}{\Lambda_\gamma^2 - m_\gamma^2} \frac{1}{4m_im_j} 
  \left( \frac{\exp (-m_\gamma r)}{r} - \frac{\exp (-\Lambda_\gamma r)}{r} 
  \right)  
  \; ; \; \gamma = \pi , K, \eta \;  .
\label{eq:vps} 
\end{eqnarray}
The $\lambda^a_{i}$ are the SU$_F$(3) flavor matrices. 
In Eq.\  (\ref{eq:vps}), experimental pseudoscalar meson 
masses $m_\pi$=138 MeV,
$m_K$=495 MeV, $m_\eta$=547 MeV, and a universal cut-off
$\Lambda_\pi$=$\Lambda_K$=$\Lambda_\eta$=4.2 fm$^{-1}$ are used. 
The quark-meson coupling constant $g_{{\rm{PS}}qq}$ is related in the usual 
way to the pion-nucleon coupling.
We furthermore include the $\sigma$-meson as the chiral partner of the
pion \cite{Wag95}, whereas we neglect the heavier (m$\simeq$1 GeV)
scalar partners of the Kaon and $\eta$.

As in Ref.\  \cite{Wag95}, we use spherical $(0s)^3$ oscillator states 
 for the baryon wave functions.
For chosen quark masses $m_u$=$m_N/3$=313 MeV and $m_u/m_s$=0.55, 
the effective quark-gluon coupling $\alpha_s$, 
the confinement strength $a_c$, and 
the wave function oscillator parameter $b_N$ 
are determined from the empirical baryon masses.
Our parameters are very similar to Ref.\ \cite{Wag95}, 
and the octet baryon magnetic moments are with the exception of the 
$\Sigma^+$ magnetic moment well described. 
%
%
Results for the individual potential contributions to the baryon masses 
are given in table \ref{table:masses}.
%

While our results for ground state hyperon properties are satisfying,
the strong one-gluon-exchange seems to prevent the simultaneous
description of the low lying Roper- and the 
negative parity resonances of the hyperons \cite{Glo96}.
The Roper resonances have been attributed to different
kinds of quark and/or meson dynamics, and their interpretation 
as pure 3-quark configurations is far from being firmly established. 
Here, we focus on electromagnetic decay amplitudes of decuplet
hyperons in order to obtain further information on the 
relative importance of pseudoscalar and vector meson exchange between
quarks.

The electromagnetic currents corresponding to the 
Hamiltonian (\ref{eq:ham}) are constructed
by a non-relativistic reduction of the Feynman diagrams 
\cite{Buc91,Buc97,Wag95} shown in Fig.\ 1.
%
%
The spatial exchange currents satisfy 
the nonrelativistic continuity equation with the exchange potentials in
Eq.\ (\ref{eq:ham}) \cite{Buc91}.
Previous quark model calculations of hyperon decays \cite{Dar83} 
were performed in impulse approximation, and only the one-body 
quark current of Fig.\ 1(a) was considered.
The PS-meson pair current ${\bf j}_{PS q\bar q}$ and 
in-flight current ${\bf j}_{\gamma PS}$ 
shown in Figs.\ 1(b) and 1(c) are given by
\begin{eqnarray}
  {\bf j}_{PS q\bar q}\! &=&\! e \left\{ 
  \exp (i{\bf q}\cdot {\bf r}_i) \b{\sigma}_i 
  \left( \b{\sigma}_j\cdot \b{\nabla}_{\bf r}\right) \left[
  \left( \b{\tau}_i\times\b{\tau}_j\right)_z {\tilde{V}}_\pi
  + \left( \lambda_4^i\lambda_5^j - \lambda_5^i\lambda_4^j \right) 
  {\tilde{V}}_K
  \right] + \left( i\leftrightarrow j\right) \right\} + 
\nonumber \\
  &+& \frac{ie}{4} \left\{ \frac{\exp (i{\bf q}\cdot {\bf r}_i)}{m_i^2} 
  \left( {\bf q} \times \b{\nabla}_{\bf r} \right)
  \left( \b{\sigma}_j\cdot \b{\nabla}_{\bf r}\right) \left[
  \left( \frac{ \b{\tau}_i\cdot\b{\tau}_j}{3}+\tau_z^{(j)} \right) 
  {\tilde{V}}_\pi +
  \frac{1}{3} \left( \lambda_4^i\lambda_4^j + \lambda_5^i\lambda_5^j -
  \right.\right.\right.
\nonumber \\
  &-& \left. \left.\left. 
  2\lambda_6^i\lambda_6^j -2 \lambda_7^i\lambda_7^j \right)  {\tilde{V}}_K 
  + \left( -\frac{1}{3} \lambda_8^i\lambda_8^j +\frac{2}{3\sqrt{3}} \lambda_8^j
  + \frac{1}{\sqrt{3}} \lambda_3^i\lambda_8^j \right) {\tilde{V}}_{\eta} 
  \right]
  + \left( i\leftrightarrow j\right) \right\}
\nonumber 
\\
  {\bf j}_{\gamma {\rm{PS}}} &=& e 
  \left( \b{\sigma}_i \cdot \b{\nabla}_{\bf r}\right) 
  \left( \b{\sigma}_j\cdot \b{\nabla}_{\bf r}\right) 
  \int_{-1/2}^{1/2} d\nu \exp \left(i{\bf q}\cdot ({\bf R} - \nu{\bf r})\right)
  \left[ \left( \b{\tau}_i\times\b{\tau}_j\right)_z  \vec{\tilde{V}}_\pi 
  + \left(
  \lambda_4^i\lambda_5^j - \lambda_5^i\lambda_4^j \right)  \vec{\tilde{V}}_K
  \right]  
\nonumber 
\\ 
  \vec{\tilde{V}}_\gamma &=& \frac{g_{{\rm{PS}} qq}^2}{4\pi} 
  \frac{\Lambda_\gamma^2}{\Lambda_\gamma^2 - m_\gamma^2} \frac{1}{4m_im_j} 
  \left( \vec{z}_{m_\gamma} \frac{\exp (-L_{m_\gamma} r)}{L_{m_\gamma} r} - 
         \vec{z}_{\Lambda_\gamma}  
  \frac{\exp (-L_{\Lambda_\gamma} r)}{L_{\Lambda_\gamma} r} \right)  
  \; ; \; \gamma = \pi , K \;  ;
\nonumber 
\\ 
  & & \vec{z}_m = L_m\vec{r} + i\nu r\vec{q}
  \quad ;\quad 
  L_m (q,\nu ,m) = \sqrt{ \frac{q^2}{4}\left( 1-4\nu^2\right) + m^2}.
\label{eq:j}  
\end{eqnarray}
In (\ref{eq:j}) ${\bf r}_i,\b{\sigma}_i,\b{\tau}_i$ are coordinate, spin and
isospin of the i-th quark, ${\bf r}={\bf r}_i-{\bf r}_j$, and ${\bf{q}}$ is the
photon momentum.
The remaining currents can be found in \cite{Wag95}.

Siegert's theorem connects the C2 and E2 transition amplitudes in the
long wavelength limit and allows to calculate the
E2 transition form factor at small momentum transfers from the charge
density $\rho ({\bf{q}})$.
 For spherical wave functions,
only the gluon- and PS-meson-pair charge density operators contribute.
\begin{eqnarray}
  \rho_{PS q\bar q} &=& \frac{ie}{2} \left\{ 
  \frac{\exp (i{\bf q}\cdot {\bf r}_i)}{m_i} 
  \left( \b{\sigma}_i\cdot {\bf q}\right)
  \left( \b{\sigma}_j\cdot \b{\nabla}_{\bf r}\right) \left[
  \left( \frac{ \b{\tau}_i\cdot\b{\tau}_j}{3}+\tau_z^{(j)} \right) 
  {\tilde{V}}_\pi +
  \frac{1}{3} \left( \lambda_4^i\lambda_4^j + \lambda_5^i\lambda_5^j -
  \right. \right. \right.   
\nonumber \\
  &-& \left. \left. \left. 
  2\lambda_6^i\lambda_6^j -2 \lambda_7^i\lambda_7^j \right) {\tilde{V}}_K
  +\left( -\frac{1}{3} \lambda_8^i\lambda_8^j +\frac{2}{3\sqrt{3}} \lambda_8^j
  + \frac{1}{\sqrt{3}} \lambda_3^i\lambda_8^j \right) {\tilde{V}}_{\eta} 
  \right]
  + \left( i\leftrightarrow j\right) \right\}
\nonumber
\\
  \rho_{gq{\bar q}} &=&
  -i {\alpha_s\over 16}\,{\b{\lambda}_i}^C\cdot{\b{\lambda}_j}^C
  {\left\{ {{e_i\over m_i} \, e^{i{\bf q}\cdot {\bf r}_i}}
  \left[  \frac{{\bf q}\cdot {\bf r}}{m_i^2} +
  \left( {{\b {\sigma}_i}\over m_i} \times {\bf{q}} \right) \cdot
  \left( {{\b {\sigma}_j}\over m_j} \times{\bf{r}} \right) \right]
  +(i\leftrightarrow j)\right\} }
  {1\over r^3}\; .
\label{eq:rho}
\end{eqnarray}
Their tensorial spin structure in Eq.(\ref{eq:rho}) allows for a double 
spin-flip of the two participating quarks 
${\b{\sigma}_i^+}{\b{\sigma}_j^-}$ as the only
mechanism by which a C2 (or E2) photon can be absorbed \cite{Buc97}.

Our M1 and C2 transition form factors are defined as
\begin{eqnarray}
  F_{\rm{M1}}({\bf{q}^2}) \!&=&\! \frac{4\sqrt{3\pi} M_N}{q} \cdot
  \langle {\rm{J}}^P{\rm{=}}\frac{3}{2}^+{\rm{, M}}_J{\rm{=}}\frac{1}{2}\,
  \vert \frac{1}{4\pi} \int d\Omega_{\bf{q}} \cdot
  Y_{\rm{1-1}}(\hat{\bf{q}}) {\bf{j}}^+({\bf{q}}) \,\vert
  {\rm{J}}^P{\rm{=}}\frac{1}{2}^+{\rm{, M}}_J{\rm{=}}\frac{1}{2} \rangle
  \; ,
\nonumber \\
  F_{\rm{C2}}({\bf{q}^2}) \!&=&\! - \frac{12\sqrt{5\pi}}{q^2} \cdot
  \langle {\rm{J}}^P{\rm{=}}\frac{3}{2}^+{\rm{, M}}_J{\rm{=}}\frac{1}{2}\,
  \vert \frac{1}{4\pi} \int d\Omega_{\bf{q}} \cdot
  Y_{20}(\hat{\bf{q}})        \rho ({\bf{q}})     \,\vert
  {\rm{J}}^P{\rm{=}}\frac{1}{2}^+{\rm{, M}}_J{\rm{=}}\frac{1}{2} \rangle \; .
\label{eq:ff}
\end{eqnarray}

\section{Results and discussion}
The one- and two-body current contributions to the M1 transition moments
$\mu$=$F_{\rm{M1}}({\bf{q}}^2{\rm{=0}})$  are given in table \ref{table:mamo}.
Individual exchange current contributions are as large as
60$\%$ of the impulse approximation result. 

As for the octet baryon magnetic moments \cite{Wag95},
we observe substantial cancellations between the
gluon-pair- and the scalar-pair-currents (confinement and one-sigma-exchange) 
 for all decays.
Due to partial cancellations between the PS-meson in-flight and the PS-meson 
pair term, the total PS-meson contribution to the M1-amplitude is small.
Nevertheless, the PS-meson contribution is important.
It reduces the strong quark-gluon coupling constant
and thus the gluon exchange current contribution.

In impulse approximation, SU$_F$(3) symmetry breaking,
i.\ e.\ the fact that $m_u/m_s=$0.55--0.6 as suggested by the octet magnetic
moments \cite{Wag95}, leads to a reduction of the transition magnetic moments
with increasing strangeness content of the hyperon.
We find that this strangeness suppression is for all six strange decays
considerably reduced when exchange currents are included.
In particular, the reduction of the
$\gamma\Xi^{0}\leftrightarrow\Xi^{\ast 0}$ M1 transition moment
$\mu_{\rm{imp}}^{\gamma\Xi^{0}\leftrightarrow\Xi^{\ast 0}}$=2.404$\mu_N$ 
with respect to the $\gamma n\leftrightarrow\Delta^0$ transition magnetic 
moment $\mu_{\rm{imp}}^{\gamma n\leftrightarrow\Delta^0}$=2.828$\mu_N$
that is observed in impulse approximation, 
practically disappears when exchange currents are included, and we obtain
$\mu_{\rm{tot}}^{\gamma n\leftrightarrow\Delta^0} \simeq
 \mu_{\rm{tot}}^{\gamma\Xi^0\leftrightarrow\Xi^{\ast 0}}$=2.428$\mu_N$.
Strangeness suppression is strong in the Skyrme model calculation of
\cite{Aba96}, while the lattice results from \cite{Lei93} agree reasonably
well with our predictions.

An interesting comparison can be made for the M1
moments of the $\gamma\Sigma^{+}\leftrightarrow\Sigma^{\ast +}$ and
$\gamma\Xi^{0}\leftrightarrow\Xi^{\ast 0}$ transitions, as well as for the 
$\gamma\Sigma^{-}\leftrightarrow\Sigma^{\ast -}$ and
$\gamma\Xi^{-}\leftrightarrow\Xi^{\ast -}$ transitions.
They are pairwise equal in impulse approximation
(see first column in table \ref{table:mamo}), and would 
also be equal after inclusion of exchange currents
if SU$_F$(3) flavor symmetry was exact.
Gluon- and scalar-exchange currents lead to 
deviations from this equality of about 10$\%$.
Less pronounced deviations from SU$_F$(3) 
(in the opposite direction) are seen in the lattice results,
whereas the Skyrme model shows a near equality for the M1 moments of the  
$\gamma\Sigma^{+}\leftrightarrow\Sigma^{\ast +}$ and
$\gamma\Xi^{0}\leftrightarrow\Xi^{\ast 0}$ transitions, but a large
difference for $\gamma\Sigma^{-}\leftrightarrow\Sigma^{\ast -}$ and
$\gamma\Xi^{-}\leftrightarrow\Xi^{\ast -}$ M1 transition.

In addition, we point out that the
transition magnetic moments for the negatively charged hyperons
($\sim$--0.4$\mu_N$) deviate considerably from the SU$_F$(3) flavor-symmetric
value 0, when the quark mass ratio $m_u/m_s$=0.55 is used.
If SU$_F$(3) symmetry was exact, these amplitudes would vanish even when
exchange currents are included.
In contrast to the Skyrme model \cite{Aba96},
we find a stronger SU$_F$(3) symmetry violation for these decays.

We observe that the hyperon transition quadrupole moments shown in
table \ref{table:quamo} receive large contributions from the 
PS-meson and gluon-pair diagrams of Fig.\ 1b and Fig.\ 1d.
We recall that the E2 transition moments resulting from the one-body 
charge and the spin-independent scalar exchange charge operators are 
exactly zero for spherical valence quark wave functions.
The transition E2 moments for the negatively charged hyperons
$\Xi^{\ast -}$ and $\Sigma^{\ast -}$ deviate from the
SU$_F$(3) flavor-symmetric value 0.
Our results are in absolute magnitude mostly larger than the Skyrme model 
results \cite{Aba96,Oh95}, 
but somewhat smaller than the lattice results \cite{Lei93}.

The helicity amplitudes
$A_{3/2}({\bf{q^2}})$ and
$A_{1/2}({\bf{q^2}})$ of the radiative hyperon decays
can be expressed as linear combinations of  the 
M1 and E2 transition form factors (\ref{eq:ff}) \cite{Gia90}.
The E2/M1 ratio of the transition amplitudes is defined as
\begin{equation}
  \frac{\rm{E2}}{\rm{M1}} \equiv \frac{1}{3}
  \frac{A_{1/2}({\rm{E2}})}{A_{1/2}({\rm{M1}})} \equiv
  - \frac{A_{3/2}({\rm{E2}})}{A_{3/2}({\rm{M1}})} =
  \frac{\omega M_N}{6} \frac{F_{\rm{C2}}({\bf{q}}^2{\rm{=0}})}
                                   {F_{\rm{M1}}({\bf{q}}^2{\rm{=0}})}
  \; .
\label{eq:e2m1}
\end{equation}
The last equality is a consequence of Siegert's theorem and has been
derived in Ref. \cite{Buc97}.
The resonance frequency $\omega$ is given in the c.m.\ system of the
decaying hyperon by
$\omega =(M_{\rm{decuplet}}^2-M_{\rm{octet}}^2)
         /(2M_{\rm{decuplet}})$.
 Following Giannini \cite{Gia90}, partial decay widths are calculated
\begin{equation}
  \Gamma_{\rm{E2,M1}} = \frac{\omega^2}{\pi}
  \frac{M_{\rm{octet}}}{M_{\rm{decuplet}}} \frac{2}{2J+1} \bigg\{
  \left\vert A_{3/2}({\rm{E2,M1}}) \right\vert^2 +
  \left\vert A_{1/2}({\rm{E2,M1}}) \right\vert^2 \bigg\} \; ; \; J=3/2
\label{eq:decaywidth}
\end{equation}
with, again,
$\vert {\rm{E2/M1}}\vert  =
 \sqrt{\Gamma_{\rm{E2}}/3\Gamma_{\rm{M1}}}$.

In table \ref{table:compar}, helicity amplitudes
$A_{3/2}({\bf{q^2}}\rm{=0})$ and $A_{1/2}({\bf{q^2}}\rm{=0})$, 
E2/M1 ratios  (\ref{eq:e2m1}) and the radiative
decay widths $\Gamma =\Gamma_{\rm{E2}}+\Gamma_{\rm{M1}}$ (\ref{eq:decaywidth})
are compared with
previous quark model calculations performed in impulse approximation \cite{Dar83},
the SU$_F$(3) Skyrme model in the slow rotor approach \cite{Aba96},
and the quenched lattice calculation of \cite{Lei93}.
Due to  cancellations of different exchange current contributions
to the M1 transition amplitude and the relative smallness of the E2 amplitude,
the decay width $\Gamma$ is dominated by the M1 impulse approximation.
This explains the agreement of the present calculation
with the results of \cite{Dar83}.
The strong suppression of the total decay width $\Gamma$ with increasing
strangeness seen in the Skyrme model is not reproduced by our calculation.
The decay widths calculated in the chiral quark model are closer 
to the lattice results.
 
The E2/M1 ratios are sensitive to exchange current contributions.
Without exchange currents and $D$-state admixtures they would all be 
identical to zero.
Except for the small $\gamma \Sigma^+\leftrightarrow \Sigma^{\ast +}$ E2/M1
ratio obtained in \cite{Aba96}, the chiral quark model, the lattice calculation
and the Skyrme model produce roughly the same ordering of E2/M1 ratios.
For five decays, the chiral quark model E2/M1 ratios lie
between the large lattice and the small Skyrme model results.
All models  yield large (the largest) E2/M1 ratios for the
negatively charged states.
However, there are important differences.
The $\gamma\Sigma^{0}\leftrightarrow\Sigma^{\ast 0}$ E2/M1 ratio in the 
Skyrme model approaches is zero \cite{Oh95,Sch95}, or almost zero \cite{Aba96},
while the SU$_F$(3) symmetry breaking and the gluon-pair current in our model 
yield a sizeable E2/M1 ratio of $-2.3\%$.
Similarly, the E2/M1 ratio for the $\Sigma^{\ast -}\rightarrow\gamma\Sigma^-$
decay, which is largely due to the gluon-pair exchange current, is with 
$-5.5\%$ almost three times larger than the Skyrme model result.
The decays of negatively charged hyperons are particularly model
dependent \cite{Sch95} due to the smallness of both the E2 and M1 
contributions. In our calculation the gluon contributes strongly to 
most E2/M1 ratios. Their measurements will give information on the 
importance of {\it effective} gluon (vector exchange) degrees of freedom 
in hadrons.  

So far experimental data exist only for the nonstrange 
$\mu^{}_{\Delta^+\rightarrow p}$ decay. We briefly summarize the
experimental situation. Most approaches (cf.\ tables \ref{table:mamo} 
and \ref{table:compar}) underestimate the empirical transition magnetic moment,
$\mu^{exp}_{\Delta^+\rightarrow p}\simeq $ 3.58(9) $\mu_N$, 
the helicity amplitudes $A^{\Delta^+\rightarrow p}_{3/2}$=--(257$\pm$8) 
                     10$^{-3}$GeV$^{-1/2}$,
                    $A^{\Delta^+\rightarrow p}_{1/2}$=--(141$\pm$5) 
                    10$^{-3}$GeV$^{-1/2}$,
and decay width $\Gamma^{exp}_{\Delta\rightarrow\gamma N}$=610--730 keV
\cite{Bar96}. Note that these observables are interrelated, and that helicity
amplitudes or decay widths are dominated by the transition magnetic moment.
According to Eq.(53) of Ref. \cite{Buc97} one obtains from the 
empirical helicity amplitudes \cite{Bar96} 
for the magnetic dipole and charge quadrupole transition form factors 
at $\vert {\bf q}\vert =0$
$F_{M1}(0)=3.58(9) \, \mu_N$ and $F_Q(0)=-0.043(40)$ fm$^2$ respectively, and
$E2/M1=-1.5(4)\%$.
Similarly, with the experimental helicity amplitudes of Ref. \cite{Bla97} 
we obtain $F_{M1}(0)=3.68(9) \, \mu_N$, $F_Q(0)=-0.105(16)$ fm$^2$, and
$E2/M1=-3.0(5)\%$, while a dispersion theoretical analysis of the Mainz data 
\cite{Bec97}
yields $F_{M1}(0)=3.47 \, \mu_N$, $F_Q(0)=-0.085(13)$ fm$^2$, and
$E2/M1=-2.5(4)\%$. Our calculated values are
$F_{M1}(0)=2.533 \, \mu_N$, $F_Q(0)=-0.089$ fm$^2$, and
 $E2/M1=-3.65\%$.
Our parameter-independent relation \cite{Buc97} between the 
transition quadrupole moment and the neutron charge radius 
$F_Q(0)=r_n^2/\sqrt{2}$ yields $F_Q(0)=-0.083$ fm$^2$, 
and the corresponding $E2$ amplitude is in very good agreement with 
recent measurements. However, the underestimation of the 
magnetic transition moment persists even after the inclusion of 
exchange currents. If we replace the calculated transition magnetic moment 
by the empirical transition magnetic moment of Ref. \cite{Bar96} 
we obtain $E2/M1=-2.6\%$ in good agreement with recent experiments.
However, this should be taken with some caution because 
experimentally one measures the {\it total} E2 and M1 amplitudes, which 
include the nonresonant Born terms, whereas we calculate only the 
{\it resonant} $N \to \Delta$ E2 and M1 amplitudes. 

\section{Summary}
We have studied the radiative decays of decuplet
hyperons within a chiral quark model including two-body exchange currents.
The present calculation complements and improves
the quark model calculations of Ref.\ \cite{Dar83}.
Exchange current effects have been evaluated for the first time for
all radiative hyperon decays.
Exchange currents have a different influence on the radiative hyperon
decay widths and on the E2/M1 ratios in these decays.
The decay width, governed by the M1-transition, is determined by
the impulse approximation because of substantial
cancellations among the various two-body currents.
Exchange currents modify the transition magnetic moments typically by
10$\%$ or less.
This is consistent with our results for octet baryon magnetic moments 
\cite{Wag95}, where similar cancellation mechanisms have been observed.
In contrast, exchange currents are extremely important for the E2/M1 ratios.
The gluon- and PS-meson-pair charge densities lead via Siegert's theorem
to non-zero E2 amplitudes for all hyperon decays. The E2/M1 ratio for the 
$\Lambda\rightarrow \gamma\Sigma^{\ast 0}$ of $-2 \%$ 
comes almost exclusively from the gluon exchange charge density. 
Experimental results on the E2/M1 ratios for the hyperon decays
provide an important test for the relative importance of 
effective gluon versus pseudoscalar degrees of freedom in low-energy QCD.

We have indicated that detailed measurements of individual M1 and E2
transition amplitudes may improve our understanding of SU$_F$(3) flavor
symmetry breaking and help to discriminate between models.
We find that strangeness suppression of the hyperon decay amplitudes is 
weaker than suggested by a recent Skyrme model calculation \cite{Aba96}.
The deviation of the decay widths of the negatively charged hyperons
$\Sigma^{\ast -}\rightarrow \gamma\Sigma^-$ and
$\Xi^{\ast -}\rightarrow \gamma\Xi^-$ 
from the SU$_F$(3) flavor symmetric value is stronger than in the Skyrme 
model. Exchange currents weaken the strangeness 
suppression observed for the transition magnetic moments 
 calculated in impulse approximation.


\begin{figure}[h]
  {\epsffile{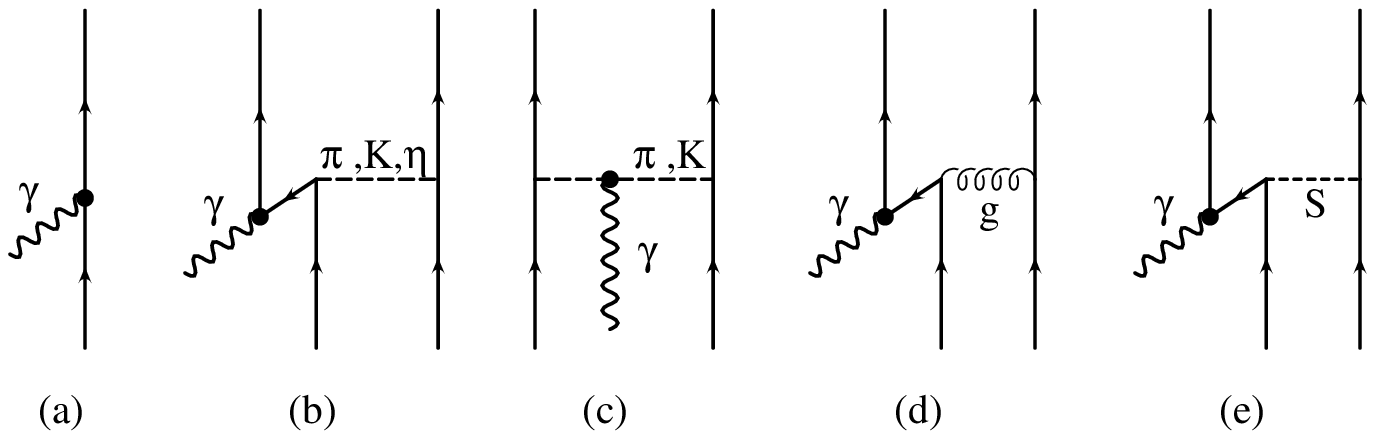}}
\caption{(a) Impulse approximation, (b) PS-meson pair current 
  ($\pi , K, \eta$),
  (c) PS-meson in-flight current, (d) gluon-pair current, and 
  (e) scalar exchange current (confinement and $\sigma$-exchange). }
\label{figure:currents}
\end{figure} 

\begin{table}[h]
\begin{center}
\begin{tabular}{ l  r  r  r  r  r  r  r | r }  
 & $\sum_i m_i$ & $E_{kin}$ & 
 $V_{conf}$ & $V_{gluon}$ & $V_{\rm{PS-octet}}$ & $V_\sigma$ & $m_B$ &
  $m_{exp}$ \cite{Bar96} \\[0.15cm] \hline 
\rule[-2mm]{0mm}{8.mm}p,n  & 
                  939 & 497 & 204 & -531 & -115 & -54 &  939 &  939 \\ 
$\Sigma$      &  1195 & 497 & 173 & -562 &  -51 & -65 & 1188 & 1193 \\ 
$\Lambda$     &  1195 & 497 & 173 & -588 &  -88 & -65 & 1124 & 1116 \\ 
$\Xi$         &  1451 & 497 & 143 & -652 &  -45 & -78 & 1316 & 1318  
\\[0.15cm] \hline 
\rule[-2mm]{0mm}{8.mm}$\Delta$ &
                  939 & 497 & 204 & -326 &  -27 & -54 & 1232 & 1232 \\
$\Sigma^\ast$ &  1195 & 497 & 173 & -423 &  -18 & -65 & 1359 & 1385 \\
$\Xi^\ast$    &  1451 & 497 & 143 & -561 &  -13 & -78 & 1439 & 1535 \\
$\Omega^-$    &  1707 & 497 & 112 & -595 &  -12 & -95 & 1615 & 1672 
\end{tabular}
\caption[Hamiltonian]{Individual 
  contributions of Hamiltonian (\ref{eq:ham}) to baryon masses. 
  All quantities are given in [MeV]. 
  Experimental values average over particles with different charge.}
\label{table:masses}
\end{center}
\end{table}

\begin{table}[h]
\begin{center}
\begin{tabular}{ l  r  r  r  r  r  r  r | c | c } 
  & ${\mu}_{\rm{imp}}$ & ${\mu}_{gq\bar q}$ &
  ${\mu}_{\rm{PS}q\bar q}$ & ${\mu}_{\gamma\rm{PS}}$ & ${\mu}_{\rm{conf}}$ &
  ${\mu}_\sigma$ & ${\mu}_{\rm{tot}}$ & 
  $\vert {\mu}_{\rm{Skyrme}}\vert$ \cite{Aba96}   &
  ${\mu}_{\rm{lattice}}$ \cite{Lei93}   
\\[0.15cm] \hline
\rule[-2mm]{0mm}{8.mm}$\gamma N\leftrightarrow \Delta$ &
   2.828 & 0.292 & -0.274 &  0.586 & -1.228 & 0.327 & 2.533 & 
   2.388 &  2.83 $\pm$ 0.49  \\
$\gamma \Sigma^+ \leftrightarrow \Sigma^{\ast +}$   & 
   2.404 & 0.366 & -0.068 &  0.097 & -0.822 & 0.291 & 2.267 & 
   1.510 &  2.22 $\pm$ 0.30  \\
$\gamma \Sigma^0 \leftrightarrow \Sigma^{\ast 0}$   & 
  -0.990 &-0.095 &  0.036 & -0.049 &  0.278 &-0.105 &-0.924 & 
   0.612 & -0.91 $\pm$ 0.11  \\
$\gamma \Sigma^- \leftrightarrow \Sigma^{\ast -}$   & 
  -0.424 &-0.176 & -0.004 &   0    &  0.267 &-0.082 &-0.419 & 
   0.286 & -0.39 $\pm$ 0.08  \\
$\gamma \Lambda \leftrightarrow \Sigma^{\ast 0}$    & 
   2.449 & 0.371 & -0.212 &  0.366 & -0.944 & 0.323 & 2.354 & 
   1.814 &  ---              \\
$\gamma \Xi^0 \leftrightarrow \Xi^{\ast 0}$         & 
   2.404 & 0.431 & -0.117 &  0.097 & -0.716 & 0.329 & 2.428 & 
   1.469 &  2.12 $\pm$ 0.24  \\
$\gamma \Xi^- \leftrightarrow \Xi^{\ast -}$         & 
  -0.424 &-0.190 &  0.009 &   0    &  0.235 &-0.090 &-0.460 & 
   0.211 & -0.36 $\pm$ 0.06 
\end{tabular} 
\caption[M1 Transition moments]{Transition 
  magnetic moments of decuplet baryons.
  The impulse (${\mu}_{imp}$) and the various exchange current contributions
  are listed separately: gluon-pair (${\mu}_{gq\bar q}$), PS-meson-pair 
  (${\mu}_{\rm{PS}q\bar q}$) and PS-meson-in-flight (${\mu}_{\gamma\rm{PS}}$),
  confinement (${\mu}_{conf}$) and $\sigma$-pair (${\mu}_{\sigma}$).
  Skyrme model results from \cite{Aba96} and 
  lattice calculation results from \cite{Lei93} (with phase conventions adapted
  to our calculation) are given in the last two columns for comparison.
  The latter are rescaled to the proton magnetic moment, which is
  too small  --  $\mu_p\simeq$ 2.3 $\mu_N$ -- on the lattice. 
  Experimentally known is only the non-strange $\Delta^+\rightarrow \gamma p$ 
  transition magnetic moment. From the empirical helicity amplitudes 
  and Eq.(53) in Ref. \cite{Buc97} one obtains      
  $\mu^{exp}_{\Delta^+\rightarrow p}=3.58(9)$ ${\mu}_N$ \cite{Bar96}, 
  $\mu^{exp}_{\Delta^+\rightarrow p}=3.68(9)$ ${\mu}_N$ 
  \cite{Bla97},  $\mu^{exp}_{\Delta^+\rightarrow p}=3.47$ ${\mu}_N$ 
  \cite{Han97}. All transition magnetic moments are given in units of 
  nuclear magnetons ${\mu}_N$= ${e\over 2M_N}$.} 
\label{table:mamo}
\end{center}
\end{table}


\begin{table}[h]
\begin{center}
\begin{tabular}{ l  c  c  c  c  c | c  c }
  & ${Q}_{gq\bar q}$ & ${Q}_{\pi q\bar q}$ &
  ${Q}_{K q\bar q}$ & ${Q}_{\eta q\bar q}$ &
  ${Q}_{tot}$ &  $\vert {Q}_{\rm{Skyrme}}\vert$ \cite{Aba96}  &
  ${Q}_{\rm{lattice}}$ \cite{Lei93} 
\\[0.15cm] \hline
  \rule[-2mm]{0mm}{8.mm}$\gamma N\leftrightarrow \Delta$ &
-0.058 & -0.027 &    0   & -0.004 & -0.089 & 0.051  &  -0.073 $\pm$ 0.190 \\
  $\gamma \Sigma^+ \leftrightarrow \Sigma^{\ast +}$ & 
-0.051 & -0.036 &  0.005 & -0.009 & -0.091 & 0.025  &  -0.141 $\pm$ 0.176 \\
  $\gamma \Sigma^0 \leftrightarrow \Sigma^{\ast 0}$ & 
 0.016 &  0.009 &  0.002 &  0.002 &  0.030 & 0.009  &   0.041 $\pm$ 0.068 \\
  $\gamma \Sigma^- \leftrightarrow \Sigma^{\ast -}$ & 
 0.018 &  0.018 & -0.010 &  0.006 &  0.032 & 0.008  &   0.050 $\pm$ 0.025 \\
  $\gamma \Lambda \leftrightarrow \Sigma^{\ast 0}$  & 
-0.041 &    0   & -0.013 &  0.006 & -0.047 & 0.035  &    ---              \\
  $\gamma \Xi^0 \leftrightarrow \Xi^{\ast 0}$       & 
-0.035 &    0   & -0.005 &  0.001 & -0.039 & 0.023  &  -0.059 $\pm$ 0.074 \\
  $\gamma \Xi^- \leftrightarrow \Xi^{\ast -}$       & 
 0.012 &    0   &  0.010 & -0.006 &  0.016 & 0.005  &   0.033 $\pm$ 0.014  
\end{tabular} 
\caption[C2 Transition moments]{Transition 
  quadrupole moments of decuplet baryons. 
  The gluon-pair (${Q}_{gq\bar q}$) and individual PS-meson 
  ($\pi ,K, \eta$) 
  exchange current contributions are listed separately. 
  Skyrme model results from \cite{Aba96} and
  lattice calculation results from \cite{Lei93} (with phase conventions adapted
  to our calculation) are given in the last two columns for comparison.
  The experimental transition quadrupole moments as extracted from 
  the empirical helicity amplitudes according to Eq.(53) in Ref. \cite{Buc97}: 
  $Q^{\rm{exp}}_{N\to \Delta}=-0.043(40)$ fm$^2$ \cite{Bar96}, 
  $Q^{\rm{exp}}_{N\to \Delta}=-0.105(16)$ fm$^2$ \cite{Bla97}, 
  $Q^{\rm{exp}}_{N\to \Delta}=-0.085(13)$ fm$^2$     \cite{Bec97,Han97}.
  All transition quadrupole moments are given in [fm$^2$]. } 
\label{table:quamo}
\end{center}
\end{table}

\begin{table}[h]
\begin{center}
\begin{tabular}{ c  r  r  r  r | r  r  | r  r | c  c } 
  & \multicolumn{4}{c|}{Chiral quark model} &
    \multicolumn{2}{c|}{Quark model  \cite{Dar83}} & 
    \multicolumn{2}{c|}{Skyrme \cite{Aba96}} &
    \multicolumn{2}{c }{Lattice \cite{Lei93}}  
\\[0.05cm]
  & $A_{3/2}$ & $A_{1/2}$ & $\Gamma$ & E2/M1 & 
  $\Gamma$ & E2/M1 & $\Gamma$ & E2/M1 & $\Gamma$ & E2/M1
\\[0.15cm] \hline
\rule[-2mm]{0mm}{8.mm}$\gamma N\leftrightarrow \Delta$ &
  -186 & -92 & 350  & -3.65& --- & --- & 309  & -2.2 & 
 430 $\pm$ 150 & 3   $\pm$ 8   \\
$\gamma \Sigma^+ \leftrightarrow \Sigma^{\ast +}$    &  
  -138 & -71 & 105  & -2.9 & 104 &  0  &  47  & -1.2 & 
 100 $\pm$  26 & 5   $\pm$ 6   \\
$\gamma \Sigma^0 \leftrightarrow \Sigma^{\ast 0}$    & 
  56   & 29  & 17.4 & -2.3 &  19 &  0  &  7.7 & -1.0 & 
 17  $\pm$ 4   & 4   $\pm$ 6   \\
$\gamma \Sigma^- \leftrightarrow \Sigma^{\ast -}$    & 
  26.1 &11.9 & 3.61 & -5.5 & 2.5 &  0  & 1.7  & -2.0 & 
 3.3 $\pm$ 1.2 & 8   $\pm$ 4   \\
$\gamma \Lambda \leftrightarrow \Sigma^{\ast 0}$     & 
  -165 & -88 & 265  & -2.0 & 232 &  0  & 158  & -1.9 & 
 ---         & ---           \\
$\gamma \Xi^0 \leftrightarrow \Xi^{\ast 0}$          & 
  -154 & -84 & 172  & -1.3 & --- & --- &  63  & -1.3 & 
 129 $\pm$  29 & 2.4 $\pm$ 2.7 \\
$\gamma \Xi^- \leftrightarrow \Xi^{\ast -}$          & 
  30   & 15  & 6.18 & -2.8 & --- & --- & 1.3  & -2.1 & 
 3.8 $\pm$ 1.2 & 7.4 $\pm$ 3.0 
\end{tabular} 
\caption[Helicity]{Helicity 
  amplitudes $A_{3/2},A_{1/2}$ (in [10$^{-3}$GeV$^{-1/2}$]),  
  radiative decay widths $\Gamma$ (in [keV])
  and E2/M1 ratios (in [$\%$]) 
  calculated in the present model in comparison with 
  impulse approximation quark model results from \cite{Dar83}, 
  SU$_F$(3) Skyrme model results (slow rotor approach for the 
  				  kaon fields) from \cite{Aba96},
  and lattice QCD results from \cite{Lei93}.
  Note that our results are given at {\bf q}$^2$=0.  
  Experimentally known are the non-strange 
  $\gamma N\leftrightarrow \Delta$ helicity amplitudes
  $A^{exp}_{3/2}$=--(257$\pm$8) 10$^{-3}$GeV$^{-1/2}$  and
  $A^{exp}_{1/2}$=--(141$\pm$5) 10$^{-3}$GeV$^{-1/2}$, and the
  decay width $\Gamma^{exp}_{\Delta\rightarrow\gamma N}$=610--730 keV
  \cite{Bar96}. 
  The empirical E2/M1 ratios for the 
  $\gamma N\leftrightarrow \Delta$ transition are 
  $E2/M1=-1.5(4)\%$ \cite{Bar96}, $E2/M1=-2.5(4)\%$ \cite{Bec97},
  $E2/M1=-3.0(5)\%$ \cite{Bla97}. If we use the empirical M1-amplitude and
  the calculated E2-amplitude we obtain $E2/M1=-2.6\%$ (see text).} 
\label{table:compar}
\end{center}
\end{table}


\begin{thebibliography}{99}
%
\bibitem{Buc91}
  A.\ Buchmann, E.\ Hern\'{a}ndez, and K.\ Yazaki,
    Phys.\ Lett.\ {\bf B269}, 35 (1991);
    Nucl.\ Phys.\ {\bf A569}, 661 (1994); 
  A.\ J.\ Buchmann,
    Z.\ Naturforsch.\ {\bf 52a}, 877 (1997).
%
\bibitem{Rob93}
  D.\ Robson,
    Nucl.\ Phys.\ {\bf A560}, 389 (1993);
  E.\ Perazzi, M.\ Radici, and S.\ Boffi,
    Nucl.\ Phys.\ {\bf A614}, 346 (1997);
%
\bibitem{Wag95}
  G.\ Wagner, A.\ J.\ Buchmann, and A.\ Faessler,
    Phys.\ Lett.\ {\bf B359}, 288 (1995).
%
\bibitem{Buc97}
  A.\ J.\ Buchmann, E.\ Hern\'{a}ndez, and A.\ Faessler,
    Phys.\ Rev.\ {\bf C55}, 448 (1997); 
  A.\ J.\ Buchmann, E.\ Hern\'{a}ndez, U.\ Meyer, and A.\ Faessler,  
    Phys.\ Rev.\ {\bf C} (1998) submitted.
%
\bibitem{Mey97}
  U.\ Meyer, A.\ J.\ Buchmann, and A.\ Faessler,
    Phys.\ Lett.\ {\bf B408}, 19 (1997).
%
\bibitem{Lip73}
  H.\ J.\ Lipkin,
    Phys.\ Rev.\ {\bf D7}, 846 (1973).
%
\bibitem{Dar83}
  J.\ W.\ Darewych, M.\ Horbatsch, and R.\ Koniuk,
    Phys.\ Rev.\ {\bf D28}, 1125 (1983);
  E.\ Kaxiras, E.\ J.\ Moniz, and M.\ Soyeur,
    Phys.\ Rev.\ {\bf D32}, 695 (1985).
%
\bibitem{Aba96}
  T.\ Haberichter, H.\ Reinhardt, N.\ N.\ Scoccola, H.\ Weigel,
    Nucl.\ Phys.\ {\bf A615}, 291 (1997);
  A.\ Abada, H.\ Weigel, H.\ Reinhardt,
    Phys.\ Lett.\ {\bf B366}, 26 (1996).
%
\bibitem{Oh95}
  Y.\ Oh,
    Mod.\ Phys.\ Lett.\ {\bf A10}, 1027 (1995).
%
\bibitem{Sch95}
  C.\ L.\ Schat, C.\ Gobbi, N.\ N.\ Scoccola,
    Phys.\ Lett.\ {\bf B356}, 1 (1995).
%
\bibitem{Kum88}
  S.\ Kumano,
    Phys.\ Lett.\ {\bf B214}, 132 (1988);
  K.\ Bermuth, D.\ Drechsel, L.\ Tiator, J.\ B.\ Seaborn,
    Phys.\ Rev.\ {\bf D37}, 89 (1988);
  I.\ Guiasu and R.\ Koniuk,
    Phys.\ Rev.\ {\bf D36}, 2757 (1987);
  M.\ Weyrauch,
    Phys.\ Rev.\ {\bf D35}, 1574 (1987);
  D.\ H.\ Lu, A.\ W.\ Thomas, A.\ G.\ Williams,
    Phys.\ Rev.\ {\bf C55}, 3108 (1997).
%
\bibitem{But93}
  M.\ N.\ Butler, M.\ J.\ Savage, and R.\ P.\ Springer,
    Phys.\ Lett.\ {\bf B304}, 353 (1993).
%
\bibitem{Lei93}
  D.\ B.\ Leinweber, T.\ Draper, R.\ M.\ Woloshyn,
    Phys.\ Rev.\ {\bf D48}, 2230 (1993).
%
\bibitem{Rus95}
  J.\ S.\ Russ,
    Nucl.\ Phys.\ {\bf A585}, 39c (1995);
  R.\ A.\ Schumacher,
    Nucl.\ Phys.\ {\bf A585}, 63c (1995).
%
%
%
\bibitem{deR75}
  A.\ DeRujula, H.\ Georgi, and S.L.\ Glashow,
    Phys.\ Rev.\ {\bf D12}, 147 (1975);
  R.\ Koniuk and N.\ Isgur,
    Phys.\ Rev.\ {\bf D21}, 1868 (1980);
  N.\ Isgur and G.\ Karl,
    Phys.\ Rev.\ {\bf D18}, 4187 (1978);
    Phys.\ Rev.\ {\bf D19}, 2653 (1979).
%
%
\bibitem{Bar96}
  Particle Data Group, R.\ M.\ Barnett et al.,
    Phys.\ Rev.\ {\bf D54}, 1 (1996).
%
\bibitem{Glo96}
  L.\ Ya.\ Glozman, Z.\ Papp, and W.\ Plessas,
    Phys.\ Lett.\ {\bf B381}, 311 (1996).
%
%
%
\bibitem{Gia90}
  M.\ M.\ Giannini,
    Rep.\ Prog.\ Phys.\ {\bf 54}, 453 (1990).
%
\bibitem{Bla97} 
  G.\ Blanpied {\it et al.}, 
    Phys.\ Rev.\ Lett.\ {\bf 79}, 4337 (1997).
%
\bibitem{Bec97} 
  R.\ Beck, {\it et al.}, 
    Phys.\ Rev.\ Lett.\ {\bf 78}, 606 (1997).
%
%
\bibitem{Han97}
  O.\ Hanstein, D.\ Drechsel and L.\ Tiator,
    Phys.\ Lett.\ {\bf B385}, 45 (1996);  see also: nucl-th/9709067.
%
\end{thebibliography}
\end{document}